\documentclass[aps,prl,preprint,showpacs]{revtex4}

\usepackage{graphicx}% Include figure files
\usepackage{dcolumn}% Align table columns on decimal point
\usepackage{bm}% bold math
\usepackage{amsfonts}
\usepackage{amsmath, amsthm, amssymb}

\newcommand{\eqz}[1]{eq.(#1)}
\newcommand{\Sch}{Schr\"odinger}
\newcommand{\im}{\mathrm{i}}		%  Imaginary Unit
\newcommand{\trace}{\mathrm{Tr}}
\newcommand{\deter}{\mathrm{det}}
\newcommand{\de}{\mathrm{d}}
\newcommand{\esp}[1]{\mathrm{e}^{#1}}
\newcommand{\fig}{Fig.}
\newcommand{\refer}{Ref.}

\begin{document}

\title{Quantum mechanics on curved 2D systems with electric and magnetic fields.}

\author{Giulio Ferrari}
\email[]{giulio.ferrari@unimore.it}
\affiliation{S3 CNR-INFM National Research Center, Via Campi 213/A, 41100 Modena, Italy.}
\author{Giampaolo Cuoghi}
\affiliation{ITIS A.Volta, Piazza Falcone e Borsellino 5, 41049 Sassuolo, Italy.}

\date{\today}

\begin{abstract}
We derive the \Sch\ equation for a spinless charged particle constrained to a curved surface with electric and magnetics fields applied.
The particle is confined on the surface using a thin-layer procedure, giving rise to the well-known geometric potential.
The electric and magnetic fields are included via the four-potential.
We find that there is no coupling between the fields and the surface curvature and that with a proper choice of the gauge, the surface and transverse dynamics are exactly separable.
Finally, the Hamiltonian for the cylinder, sphere and torus are analytically derived.
\end{abstract}

%\keywords{}

\maketitle
Two dimensional (2D) curved systems are extensively investigated to study new physical effects that depend both on the curvature of the systems and on the external electric and magnetic fields applied, such as Aharonov-Bohm effect \cite{Bachtold99}, formation of Landau levels \cite{Aoki92,Kim92} and quantum Hall effect \cite{Perfetto07}.
Nanostructures with a great variety of novel geometries are now experimentally produced. At the same time, sources of high magnetic fields are accessible. Hence, a rigorous theoretical understanding of the dynamics under such condition is needed.
Dynamics on curved surfaces has become particularly important in condensed matter since the synthesis of curved graphene systems, such as fullerenes and carbon nanotubes. The fullerenes may show effects induced by the magnetic field on the photocurrent for an intensity below 1 T \cite{Osipyan99}. The carbon nanotube radius is usually too small to allow for significant effects induced by experimentally accessible magnetic fields. However field-induced effects may become important in multiwalled carbon nanotubes, where the radius is of the order of some tens of nanometers and a field of tens of Tesla is sufficient to see effects on the energy band gap \cite{Coskun04,Lassagne07}. New techniques have also been developed to obtain semiconductor tubes that have a radius ranging from tens of nanometers up to microns. With such dimensions, fields weaker that 10 T can show significant effects on the magnetoresistance \cite{Shaji07, Vorobev07}.
These successes on the experimental side push for a theoretical comprehension of the quantum carrier mechanics on curved structures immersed in magnetic fields. Historically, two methods have been employed to study curved systems: a method due to DeWitt \cite{DeWitt57} that approaches the problem by studying the dynamics as fully 2D and another due to da Costa \cite{daCosta81} that derives the \Sch\ equation starting from the three dimensional (3D) one and then reduces it to a 2D equation by a confining procedure.
If no magnetic field is applied the procedure of da Costa is widely used and accepted \cite{Encinosa03,Marchi05}. This procedure appears to be the most rigorous and physically sound for curved nanostructures, since a DeWitt-like 2D Lagrangian approach does not allow for the inclusion of an arbitrarily oriented 3D magnetic field but only perpendicular to the surface. Moreover, these structures are 2D systems embedded in a 3D space, and the da Costa approach describes exactly this situation. In spite of these considerations, a rigorous approach has not been completely developed including the magnetic field. For example, studying cylindrical geometries only a magnetic field perpendicular to the surface has been considered effective for the dynamics \cite{Ando05,Perfetto07,Vorobev07}; also for toroidal surfaces the same approach has generally been adopted \cite{Onofri01,Encinosa05}; while the simple geometry of the sphere does not allow to distinguish between the different approaches \cite{Aoki92,Kim92,Pudlak07}. Nevertheless, the da Costa method is recognised as the one to be employed \cite{Encinosa06}, but an analytical expression for the \Sch\ equation including the magnetic field has not been derived yet.

In this paper, we follow the procedure of da Costa including the effect of the magnetic field via the vector potential $\mathbf{A}$ and the electric field via the scalar potential $V$. We shall derive analytically a \Sch\ equation valid for any 2D geometry, that describes in the most appropriate way real curved nanostructures with electric and magnetic field applied, given the above considerations. We shall show that there is no coupling between the field and the surface curvature and that the dynamics on the surface is decoupled from the transverse one with a proper choice of the gauge, without approximations.

Here and in the following $i,j,k$ stand for the spatial indices and assume the values $1,2,3$. Tensor covariant and contravariant components are used and Einstein summation convention is adopted.
We define the gauge covariant derivative
$D_j=\nabla_{j}-\frac{\im Q}{\hbar} A_j$,
where $Q$ is the charge of the particle and $A_j$ the covariant components of the vector potential $\mathbf{A}$.
The covariant derivative $\nabla_{j}$ is defined as
$\nabla_{j} v^i=\partial_{j}v^i+\Gamma^{i}_{jk}v^k$,
where $v^i$ are the contravariant components of a 3D vector field $\mathbf{v}$, $\Gamma^{i}_{jk}$ are the Christoffel symbols and $\partial_{j}$ is the derivative with respect to the spatial variable $q_j$.
The covariant 3D \Sch\ equation, containing both the vector potential and the electric potential, is
\begin{equation}
\label{eq:schr1}
\im\hbar\frac{\partial}{\partial t}\psi=-\frac{\hbar^2}{2m}G^{ij}D_i D_j\psi+Q V \psi,
\end{equation}
where the metric tensor $G_{ij}$, and its inverse $G^{ij}$, has been introduced to take into account the geometry of the space.
Defining the scalar potential $A_0=-V$, we can define a gauge covariant derivative for the time variable as
$D_0={\partial_t}-{\im Q} A_0/\hbar$,
and rewrite \eqz{\ref{eq:schr1}} as
\begin{equation}
\label{eq:schr2}
\im\hbar D_0 \psi=-\frac{\hbar^2}{2m}G^{ij}D_i D_j\psi.
\end{equation}
The gauge invariance of the above equation can be easily demonstrated with respect of the following gauge transformations:
\begin{eqnarray}
\label{eq:gaugetrasf}
A_j \to A'_j=A_j+\partial_j \gamma\; ;\;A_0 \to A'_0=A_0+{\partial_t} \gamma;\nonumber\\
\\
\psi \to \psi'=\psi\esp{\im Q \gamma/\hbar}\nonumber,
\end{eqnarray}
where $\gamma$ is a scalar function.
We expand \eqz{\ref{eq:schr2}} by covariant calculus, obtaining
\begin{equation}
\label{eq:schr3}
\im\hbar D_0 \psi=
\frac{1}{2m}
\left[
-\frac{\hbar^2}{\sqrt{G}}\partial_i\left(\sqrt{G} G^{ij}\partial_j\psi\right)
+\frac{\im Q \hbar}{\sqrt{G}}\partial_i\left(\sqrt{G} G^{ij} A_j\right)\psi
+2\im Q \hbar G^{ij} A_j \partial_i \psi
+Q^2 G^{ij} A_i A_j \psi
\right],
\end{equation}
where $G=\deter(G_{ij})$.
The above equation is the covariant \Sch\ equation for a generic 3D curvilinear coordinate system, when electric and magnetic fields are applied.
Note that no gauge has been chosen, but the general expression $\mathbf{A}=(A_1,A_2,A_3)$ valid for any gauge and any magnetic field will be maintained through the paper until differently stated.

Before applying the thin-layer procedure described by da Costa \cite{daCosta81} to confine the particle on the surface, the coordinate system has to be described. The system description is analogous to the one given in \refer\ \cite{daCosta81}.
The surface $S$ is parametrised by $\mathbf{r}=\mathbf{r}(q_1,q_2)$,where $\mathbf{r}$ is the position vector of an arbitrary point on the surface.
The 3D space in the immediate neighbourhood of $S$ can be parametrised as
$\mathbf{R}(q_1,q_2,q_3)=\mathbf{r}(q_1,q_2)+q_3\mathbf{n}(q_1,q_2)$,
where $\mathbf{n}(q_1,q_2)$ is the unit vector normal to $S$.
For the sake of clarity, we introduce the indices $a,b$ to indicate the surface parameters, which hence assume the values $1,2$.
The relation between the 3D metric tensor $G_{ij}$ and the 2D induced one $g_{ab}={\partial_a \vec{r}}\cdot{\partial_b \vec{r}}$ is:
\begin{eqnarray}
\label{eq:metric}
G_{ab}=g_{ab}+\left[\alpha g+(\alpha g)^T\right]_{ab}q_3+(\alpha g \alpha^T)_{ab}q_3^2\nonumber\\
\\
G_{a3}=G_{3a}=0, \; G_{33}=1\nonumber,
\end{eqnarray}
where $\alpha_{ab}$ is the Weingarten curvature matrix for the surface \cite{daCosta81,Wolfram}.
The structure of the metric tensor given in \eqz{\ref{eq:metric}} suggests to separate the \Sch\ \eqz{\ref{eq:schr3}} in a surface part for $a,b=1,2$ and a normal part.
Besides, a confining potential $V_{\lambda}(q_3)$ is assumed to localise the particle on the surface $S$, where $\lambda$ is a parameter which measures the strength of the confinement. We follow a well-established thin-layer method \cite{daCosta81,Encinosa06}. Since the aim of the procedure is to obtain a surface wave-function depending only on $(q_1,q_2)$, we introduce a new wave-function
$
\chi(q_1,q_2,q_3)=\chi_{S}(q_1,q_2) \chi_{n}(q_3).
$
The separability is an hypothesis and shall be verified.
The condition of conservation of the norm gives the relation:
\begin{equation}
\label{eq:chichi}
\psi(q_1,q_2,q_3)=\left[ 1+\trace(\alpha)q_3+\deter(\alpha)q_3^2 \right]^{-1/2}\chi(q_1,q_2,q_3).
\end{equation}
First, we substitute expression (\ref{eq:chichi}) into \eqz{\ref{eq:schr3}}. Then we take into account the effect of the potential $V_{\lambda}(q_3)$: in the limit of confinement, the wave-function is localised on $S$ by two step potential barriers on both sides of the surface. This means that the value of the wave-function is different from zero only in a close neighbourhood of $S$. We can thus perform the limit $q_3\to0$ in the \Sch\ equation. The final result is
\begin{eqnarray}
\label{eq:shro2D}
\im\hbar D_0 \chi&=&
\frac{1}{2m}
\left[
-\frac{\hbar^2}{\sqrt{g}}\partial_a\left(\sqrt{g}g^{ab}\partial_b \chi\right)
+\frac{\im Q \hbar}{\sqrt{g}}\partial_a\left(\sqrt{g}g^{ab} A_b\right)\chi
+2\im Q \hbar g^{ab} A_a \partial_b \chi\right.+ \nonumber\\
&& +Q^2 \left(g^{ab}A_aA_b+(A_3)^2\right)\chi
-\hbar^2\left(\partial_3\right)^2\chi
+{\im Q \hbar}\left(\partial_3 A_3\right)\chi
+2\im Q \hbar A_3 \left(\partial_3 \chi\right)+\\
&&\left.-\hbar^2\left(\left[\frac{1}{2}\trace(\alpha)\right]^2-\deter(\alpha)\right)\chi
\right]+V_{\lambda}(q_3)\chi,\nonumber
\end{eqnarray}
where $g=\deter{(g_{ab})}$ and all the components of $\mathbf{A}$ and its derivative are calculated at $q_3=0$.
From the above equation, we can state the first fundamental evidence of this paper: {\em There is no coupling between the magnetic field and the curvature of the surface, independently of the shape of the surface, of the field $\mathbf{B}$ and of the gauge}. In fact, in the equation \eqz{\ref{eq:shro2D}} terms mixing $A_j$ and the curvature matrix $\alpha_{ab}$ do not appear. This is in contrast with what obtained in Ref.\cite{Encinosa06}, where the apparent coupling between the field and the curvature is due to the choice of a particular gauge in the derivation of the formula.

Note that in \eqz{\ref{eq:shro2D}} the well-known geometric potential $V_{S}$ appears \cite{daCosta81}:
\begin{equation}
V_{S}(q_1,q_2)=-\frac{\hbar^2}{2m}\left(\left[\frac{1}{2}\trace(\alpha)\right]^2-\deter(\alpha)\right),
\end{equation}
where the first term is the square of the mean curvature and the second one is the Gaussian curvature.

We next verify that the limiting procedure preserves the gauge invariance of the resulting equation. Defining a new metric tensor $\tilde{G}$ as
\begin{equation}
\tilde{G}=
\left(
\begin{array}{ccc}
g_{11}&g_{12}&0\\
g_{21}&g_{22}&0\\
0&0&1
\end{array}
\right),
\end{equation}
\eqz{\ref{eq:shro2D}} can be rewritten in a compact form:
\begin{equation}
\label{eq:shrogauge}
\im\hbar D_0 \chi=
-\frac{\hbar^2}{2m}\tilde{G}^{ij}\tilde{D}_i \tilde{D}_j\chi+V_{S}\chi+V_{\lambda}(q_3)\chi,
\end{equation}
so that the invariance with respect to the gauge transformations (\ref{eq:gaugetrasf}) is evident in the above expression.

Finally, we demonstrate the separability of the dynamics on the surface and perpendicular to the surface, that is our work hypothesis. In \eqz{\ref{eq:shro2D}} only one term, $\left(A_3(q_1,q_2,0)\partial_3 \chi\right)$, couples the dynamics along $q_3$ with the dynamics on $S$. Since we have shown the gauge invariance of \eqz{\ref{eq:shrogauge}}, we can now impose a gauge such to cancel the component $A_3$ of the vector potential, cancelling the coupling term.
Applying the gauge transformations (\ref{eq:gaugetrasf}), the best suitable choice for $\gamma$ is
\begin{equation}
\label{eq:ggauge}
\gamma(q_1,q_2,q_3)=-\int_0^{q_3} A_3(q_1,q_2,z) \de z .
\end{equation}
We obtain
$A'_3=0, \; \partial_3 A'_3=0$ and
having fixed the lower limit of integration to 0, in the limit $q_3 \to 0$, $A_1$ and $A_2$ remain unchanged.
After the gauge transformation we can separate the \Sch\ equation in two independent equations: 
\begin{equation}
\label{eq:dynq3}
\im \hbar \partial_t \chi_{n}=-\frac{\hbar^2}{2m}(\partial_3)^2\chi_{n}+V_{\lambda}(q_3)\chi_{n},
\end{equation}
\begin{eqnarray}
\label{eq:shrodecop}
\im\hbar \partial_t \chi_{S}&=&
\frac{1}{2m}
\left[
-\frac{\hbar^2}{\sqrt{g}}\partial_a\left(\sqrt{g}g^{ab}\partial_b \chi_{S}\right)
+\frac{\im Q \hbar}{\sqrt{g}}\partial_a\left(\sqrt{g}g^{ab} A_b\right)\chi_{S}
+2\im Q \hbar g^{ab} A_a \partial_b \chi_{S}
\right. +\nonumber \\
&& \left. +Q^2g^{ab}A_aA_b \chi_{S}\right]
+V_{S}\chi_{S}
+Q V\chi_{S},
\end{eqnarray}
where we have made explicit both the confining potential and the electric potential. Expression (\ref{eq:dynq3}) is the 1D \Sch\ equation for a particle confined by the potential $V_{\lambda}$, while expression (\ref{eq:shrodecop}) describes the dynamics of a particle bounded to the surface under the effect of electric and magnetic fields. Note that the separation of the dynamics has been obtained analytically without any approximation. At this point we can state the second fundamental conclusion of this paper: {\em With a proper choice of the gauge, the dynamics on the surface and the transverse dynamics are decoupled}. In Ref.\cite{Encinosa06} this separability is not evident because of the not-optimal choice of the gauge.

\begin{figure*}
\centering
\includegraphics[width=15cm, angle=0]{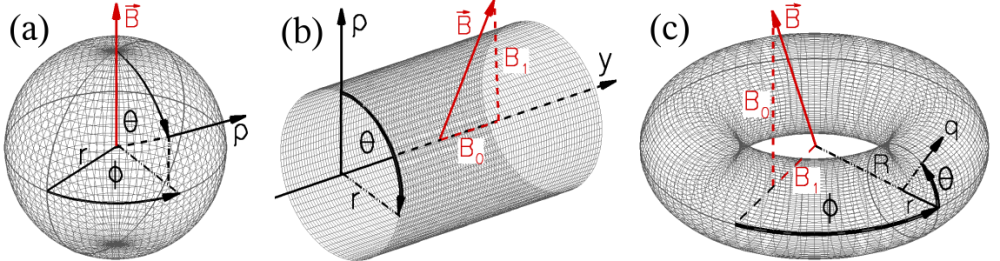}
\caption{\label{fig:superfici} (a) A spherical surface  of radius $r$ and its coordinate system $({\theta},{\phi},{\rho})$. The coordinate system is chosen so that the magnetic field $\mathbf{B}$ is along the polar direction. (b) A cylindrical surface  of radius $r$ and its coordinate system $({\theta},{y},{\rho})$. The magnetic field $\mathbf{B}$ and its component $B_0$ parallel to the axis and $B_1$ perpendicular to the axis at $\theta=0$ are shown. (c) A toroidal surface and its coordinate system $({\theta},{\phi},{q})$. $R$ is the distance from the centre of the tube to the centre of the torus, $r$ is the radius of the tube. The magnetic field $\mathbf{B}$ and its component $B_0$ perpendicular to the torus plane and $B_1$ laying in the torus plane are shown.}
\end{figure*}

In the following we give some examples of curved surfaces of typical nanostructures, with a homogeneous magnetic field applied.
The sphere is the simplest curved geometry to be investigated.
Given a sphere of radius $r$ and a constant magnetic field $\mathbf{B}$ in a given direction, the spherical coordinate system $({\theta},{\phi},{\rho})$ is set with the polar axis along the direction of the field, as shown in panel (a) of \fig\ref{fig:superfici}.
The most suitable vector potential, determined by the gauge condition (\ref{eq:ggauge}) for the spherical geometry is
$(A_{\theta},A_{\phi},A_{\rho})=\left(0,\frac{1}{2}B{r^2\sin^2\theta},0\right)$ and
the corresponding \Sch\ equation is
\begin{equation}
\im\hbar \partial_t \chi_{S}=
\frac{1}{2m}
\left[
-\frac{\hbar^2}{r^2}\left(\frac{1}{\sin\theta}\partial_{\theta}\left(\sin\theta\partial_{\theta}\chi_{S}\right)+\frac{1}{\sin^2 \theta}\partial^2_{\phi}\chi_{S}\right)
+\im Q \hbar B \partial_{\phi} \chi_{S}
+\frac{1}{4} Q^2 B^2 r^2 \sin^2\theta\chi_{S}
\right].
\end{equation}
Note that for a sphere $V_{S}=0$.
In literature, a number of papers appear investigating the effect of a magnetic field applied to a sphere \cite{Osipyan99,Aoki92,Kim92,Pudlak07}. Given the simple geometry of the sphere, the \Sch\ equation employed in those paper has the correct form.

A very popular geometry is the cylindrical one, given the extensive investigation on carbon nanotubes and semiconductor nanotubes \cite{Bachtold99,Ando05,Shaji07,Vorobev07,Perfetto07,Lassagne07}.
Given a cylindrical coordinate system $({\theta},{y},{\rho})$, a field $\mathbf{B}$ applied to a cylinder of radius $r$ can always be decomposed in a component $B_0$ parallel to the axis and a component $B_1$ perpendicular to the axis at $\theta=0$. The system is shown in panel (b) of \fig\ref{fig:superfici}.
The proper vector potential determined by \eqz{\ref{eq:ggauge}} is
$(A_{\theta},A_{y},A_{\rho})=\left(\frac{1}{2}r^2 B_0,rB_1 \sin\theta,0\right)$.
We can then calculate the \Sch\ equation
\begin{eqnarray}
\im\hbar \partial_t \chi_{S}&=&
\frac{1}{2m}
\left[
{-\hbar^2}\left(\frac{1}{r^2}\partial^2_{\theta}\chi_{S}+\partial^2_{y}\chi_{S}\right)
+\im Q \hbar B_0 \partial_{\theta} \chi_{S} +2\im Q \hbar r B_1 \sin\theta\partial_{y} \chi_{S} \right. +\nonumber\\
&&\left.
+Q^2 r^2\left(\frac{1}{4} B_0^2+B_1^2\sin^2\theta\right) \chi_{S}
\right]
-\frac{\hbar^2}{8m r^2}\chi_{S}.
\end{eqnarray}
Several theoretical studies on the effect of the magnetic field applied to 2D cylindrical systems have been carried out \cite{Perfetto07,Ando05}. The most widely used procedure to address the problem is to write the \Sch\ equation obtained with the 2D Laplacian generalised including the 2D vector potential: the result is not rigorous, because it does not take into account the effect of the component $B_0$. Also the surface potential $V_{S}$ is not obtained, but this reduces only to a constant shift in energy, if the radius of the tube is constant and no bending is considered \cite{Marchi05}.

A toroidal surface is more interesting both from the theoretical and experimental point of view: for example localisations are predicted also without applied magnetic field \cite{Encinosa05}, also band-gap modulations are expected \cite{Rocha04}. The particular topology is a test-bed for models on curved surfaces \cite{Onofri01,Encinosa06}.
The reference system $({\theta},{\phi},{q})$, described in panel (c) of \fig\ref{fig:superfici}, can always be chosen so that a field in an arbitrary direction can be described with a component $B_1$ in the torus plane and a component $B_0$ perpendicular to the torus plane.
$R$ is the distance from the centre of the tube to the centre of the torus, $r$ is the radius of the tube.
Using \eqz{\ref{eq:ggauge}}, we can calculate the vector potential most suitable for a toroidal surface, that is
$(A_{\theta},A_{\phi},A_{q})=\left(\frac{1}{2}B_1r\sin\phi (R\cos\theta+r),\frac{1}{2}W(\theta)\left(B_0 W(\theta)-B_1 r\sin\theta \cos\phi\right),0\right)$,
where $W(\theta)=R+r\cos\theta$. The \Sch\ equation is obtained from \eqz{\ref{eq:shrodecop}}:
\begin{eqnarray}
\im\hbar \partial_t\chi_{S} &=&
\frac{1}{2m}\left\{-\frac{\hbar^2}{r^{2}}\partial_{\theta}^{2}\chi_{S}+\frac{\hbar^2\sin\theta}{rW(\theta)}\partial_{\theta}\chi_{S}-\frac{\hbar^2}{W^{2}(\theta)}\partial_{\phi}^{2}\chi_{S}-\left(\frac{\hbar R}{2rW(\theta)}\right)^2\chi_{S}\right. + \nonumber\\
&&+\frac{iQ\hbar B_{1}\sin\phi(R\cos\theta+r)}{r}\partial_{\theta}\chi_{S}+\frac{iQ\hbar\left(B_{0}W(\theta)-B_{1}r\sin\theta\cos\phi\right)}{W(\theta)}\partial_{\phi}\chi_{S} + \\
&&-iQ\hbar B_{1}\sin\theta\sin\phi \frac{R^2+2rR\cos\theta}{2rW(\theta)} \chi_{S}
+\frac{Q^{2}}{4}\left[\left(B_{1}W(\theta)\sin\phi\right)^{2}+\left(B_{0}W(\theta)\right)^{2} \right. + \nonumber\\
&&+\left.\left.\left(B_{1}r\sin\theta\right)^{2}-2B_{0}B_{1}rW(\theta)\sin\theta\cos\phi
-(B_{1}R\sin\theta\sin\phi)^2\right]\chi_{S}\right\}. \nonumber
\end{eqnarray}
It is important to note that the above expression contains all the terms influencing the dynamics and it has been obtained exactly without any approximation. Differences can be noted with respect to the corresponding expression in \refer\cite{Encinosa06}: the discrepancies are to be attributed to the approximations performed in that paper.

In conclusion, we have rigorously developed a general \Sch\ equation valid for any 2D curved structure when magnetic and electric fields are applied. We have shown that there is no coupling between the surface curvature and the magnetic field. Moreover, we have demonstrated that with a proper choice of the gauge, the dynamics on the surface is analytically decoupled from the transverse one. To show the effectiveness of the method, we have calculated analytically the complete \Sch\ equation for a charged particle bounded to the surface of a sphere, of a cylinder and of a torus, with a homogeneous magnetic field applied in an arbitrary direction.

Acknowledgements:
This work has been partially supported by project FIRB-RBIN04EY74.
The authors thank F. Bastianelli, A. Bertoni and G. Goldoni for the helpful discussions.

\bibliography{ferrari}

\end{document}